\documentclass[aps,pra,showpacs,twocolumn,byrevtex]{revtex4}
\usepackage{graphicx}

\begin{document}

\title{Is the phase of plane waves a frame-independent quantity?}

\author{Aleksandar Gjurchinovski}

\email{agjurcin@iunona.pmf.ukim.edu.mk}

\affiliation{Department of Physics, Faculty of Natural Sciences 
and Mathematics, Sts.\ Cyril and Methodius University,
P.\ O.\ Box 162, 1000 Skopje, Macedonia}

\begin{abstract}

The invariance of the phase of plane waves among inertial frames is investigated
in some details. 
The reason that eventually led the author of a recent EPL letter [EPL \textbf{79}, 1006 (2007)]
to a spurious conclusion of the non-invariance of the phase of waves has been
identified -- it is the ignorance of the effect of relativistically-induced 
optical anisotropy in the analysis of the problem.
It is argued that the Lorentz-invariant expression for the phase of waves should be taken in the 
form $\Phi=\mathbf{k\cdot r}-\mathbf{k\cdot u}/c$, instead of the widely-used expression 
$\Phi=\mathbf{k\cdot r}-\omega t$ which has a limited validity.

\end{abstract}


\date{January 24, 2008}

\maketitle

\section{Introduction}

In a recent EPL letter by Huang \cite{huang}, the invariance of the phase of plane waves among inertial frames
has been challenged for the case of a ``superluminal'' motion of the medium, 
when the wave propagation in a stationary optical medium is observed from a frame 
traveling in the direction of the wave at a speed larger than the speed of the wave with 
respect to the medium. Apparently, by considering the phase of waves 
$\Phi=\mathbf{k\cdot r}-\omega t$ to be a Lorentz scalar, Huang obtained negative
values for the frequency of the wave with respect to the frame in which the medium is
in motion. To overcome this difficulty, Huang
introduced two types of relativistic transformation for the four-vector
$k^\mu=(\mathbf{k},\omega/c)$ based on differential Lorentz transformation. 

In this letter, we argue that the reason that eventually led Huang to this apparent non-invariance of the phase
of waves is the ignorance of the effect of relativistically-induced optical anisotropy 
\cite{bolotovskii,gjurchinovski1,gjurchinovski2} in the analysis 
of the problem. In particular, the scalar (dot) product between the wave vector $\mathbf{k}$ (the wavefront normal) 
and the velocity $\mathbf{u}$ of the wave may have different signs with respect to different inertial 
reference frames. We conclude that the phase invariance among inertial frames is 
preserved if the expression for the phase is taken in the form $\Phi=\mathbf{k\cdot r}-\mathbf{k\cdot u}/c$.

\section{Which expression for the phase is a Lorentz-invariant?}

Let us consider a three-dimensional plane-wave disturbance with a harmonic wave 
form \cite{hecht,born,jackson}. We choose the profile of the wave to vary in space sinusoidally, 
namely $\psi(\mathbf{r},t=0)=A\sin(\mathbf{k\cdot r})$. We observe that 
$\psi(\mathbf{r},t=0)$ is constant when $\mathbf{k\cdot r}=const$, which is an equation 
defining a set of planes perpendicular to the vector $\mathbf{k}$. The value of
$\psi(\mathbf{r},t=0)$ repeats itself in space after a displacement $n\lambda$ in the
direction of $\mathbf{k}$, where $\lambda$ is the wavelength of the wave, and $n$ is an integer. 
\begin{figure}
\includegraphics[width=.40\textwidth,height=!]{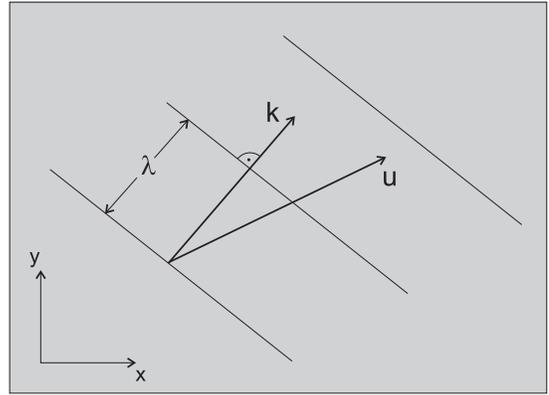}
\caption{A two-dimensional sketch of a plane wave traveling at a velocity $\mathbf{u}$.
The wave vector $\mathbf{k}$ is normal to the wavefronts, but not necessarily parallel
to the velocity $\mathbf{u}$ of the wave.}
\end{figure} 
If the profile of the wave disturbance moves at a velocity
$\mathbf{u}$ (see Fig. 1), it can be described by a harmonic wave function in the form
\begin{equation}
\psi(\mathbf{r},t)=A\sin\mathbf{k}\cdot(\mathbf{r}-\mathbf{u}t)=A\sin\Phi,
\end{equation}
which we obtain by merely replacing $\mathbf{r}$ in $\psi(\mathbf{r},t=0)$ with 
$\mathbf{r}-\mathbf{u}t$. The expression: 
\begin{equation}
\Phi=\mathbf{k\cdot r}-\mathbf{k\cdot u}t
\end{equation}
is the phase of the wave, 
and $\mathbf{k}$ is the wave vector, having a magnitude $2\pi/\lambda$ and pointing in the 
direction orthogonal to the planes $\Phi=const$. For a fixed stationary observer at 
$\mathbf{r=r_0}$ the wave disturbance $\psi(\mathbf{r_0},t)$ repeats itself in time after 
a temporal period $T=2\pi/|\mathbf{k\cdot u}|$, the inverse of which is the frequency $f$ of the wave.
The quantity $|\mathbf{k\cdot u}|$, which is the absolute value of the dot product between 
the wave vector and the velocity of the wave,
is referred to as the angular frequency of the wave, commonly
denoted by $\omega$. In the case of isotropic media (including vacuum), 
the wave vector $\mathbf{k}$ and the velocity $\mathbf{u}$ of the wave are parallel and 
pointing in the same direction, allowing the expression for the phase of the wave in Eq. (2) 
to be recasted in the following widely-used form:
\begin{equation}
\Phi=\mathbf{k\cdot r}-\omega t.
\end{equation}

If the plane-wave disturbance in Eq. (1) is observed from a different reference frame, 
the phase of the wave $\Phi$ should remain invariant quantity. This claim is clarified
by the fact that the elapsed phase of the wave is proportional to the number of wavecrests
that have passed the observer, and thus it must be frame-independent, and hence, a Lorentz scalar
\cite{jackson,rindler}.
Alternatively, the same conclusion follows by considering optical interference experiments from 
different inertial frames, where the phase $\Phi$ is the quantity that determines the 
interference pattern \cite{stephani}. 

In the aforementioned letter to EPL \cite{huang}, Huang used the invariance of the phase $\Phi$ in Eq. (3)
to Lorentz-transform the wave characteristics between different inertial frames. In this way, Huang obtained
negative angular frequencies in the frame where the medium moves at ``superluminal'' speeds against the wave. 
What has not been taken into account in Huang's analysis is that the dot product between 
the wave vector and the velocity of the wave changes its sign from positive to negative
when switching between the medium's rest frame and the frame in which the medium moves ``superluminally''.
Let us clarify this point more explicitly.
The approach by using Eq. (3) as an expression for the phase 
will work efficiently if the wave is propagating in vacuum. In the case of vacuum, the wave 
vector and the velocity of the wave will remain parallel and unidirectional with respect to 
any inertial frame. Hence,
$\mathbf{k\cdot u}=|\mathbf{k\cdot u}|=\omega$ and $\mathbf{k'\cdot u'}=|\mathbf{k'\cdot u'}|=\omega'$, and the phase 
invariance $\mathbf{k\cdot r}-\mathbf{k\cdot u}t=\mathbf{k'\cdot r'}-\mathbf{k'\cdot u'}t'$ between the frames 
$S$ and $S'$ would imply $\mathbf{k\cdot r}-\omega t=\mathbf{k'\cdot r'}-\omega' t'$.
However, when the wave propagates in a material medium, the invariance of the phase given in
Eq. (3) will generally not work even if the optical medium is isotropic in its rest frame. 
This is due to the fact that an optical medium that is optically isotropic in its rest frame $S'$
will possess an optical anisotropy in the frame $S$ in which it is moving at a constant velocity.
It should be noted that this induced optical anisotropy is of a purely relativistic
origin, and it has nothing to do with the usual anisotropy in the crystals \cite{gjurchinovski2}.
Hence, while the wave vector and the velocity of the wave are parallel and unidirectional 
with respect to the rest frame $S'$ of the medium, this may not be the case with respect
to some reference frame $S$ in which the medium is in motion. 
In fact, it might happen that the dot product between the wave vector (the wavefront normal) 
and the velocity of the wave may have different signs with respect to different reference frames.
In particular, the phase invariance $\mathbf{k\cdot r}-\mathbf{k\cdot u}t=\mathbf{k'\cdot r'}-\mathbf{k'\cdot u'}t'$
may not imply $\mathbf{k\cdot r}-\omega t=\mathbf{k'\cdot r'}-\omega' t'$ in the latter case, since it may 
happen that $\mathbf{k\cdot u}=-|\mathbf{k\cdot u}|=-\omega<0$ when $\mathbf{k'\cdot u'}=\mathbf{|k'\cdot u'|}=\omega'>0$.
In this sense, the expression for the phase of the wave in Eq. (2) is more general (and thus, 
more correct) than the one in Eq. (3), and therefore, it is the expression 
$\Phi=\mathbf{k\cdot r}-\mathbf{k\cdot u}t$ that should have a role of Lorentz scalar between different 
inertial reference frames.

\section{Relativistic transformations of the wave characteristics}

Consider from a frame $S$ a plane monochromatic wave propagating at a velocity $\mathbf{u}$,
whose wavefront normal is described by the wave vector $\mathbf{k}$. With respect to 
the frame $S'$ which moves at a constant velocity $\mathbf{V}$ relative to $S$, the velocity of
the wave is $\mathbf{u'}$, and $\mathbf{k'}$ is the corresponding wavefront normal. We assume
that the corresponding axes of $S'$ and $S$ are parallel, and that the velocity $\mathbf{V}$ of $S'$ in
$S$ is in an arbitrary direction. The invariance of the phase implies:
\begin{equation}
\mathbf{k\cdot r}-\mathbf{k\cdot u}t=\mathbf{k'\cdot r'}-\mathbf{k'\cdot u'}t'.
\end{equation}
By applying the Lorentz transformation between $S$ and $S'$ \cite{jackson}:
\begin{eqnarray}
\mathbf{r'}&=&\mathbf{r}+(\gamma-1){(\mathbf{r\cdot V})\over \|\mathbf{V}\|^2}\mathbf{V}-\gamma\mathbf{V}t,\\
t'&=&\gamma\left(t-{\mathbf{r\cdot V}\over c^2}\right),\\
\gamma&=&\left(1-{\|\mathbf{V}\|^2\over c^2}\right)^{-1/2},
\end{eqnarray}
into Eq. (4), we obtain:
\begin{eqnarray}
\mathbf{k\cdot r}&-&\mathbf{k\cdot u}t=\nonumber\\
&=&\left[\mathbf{k'}+\mathbf{V}\left((\gamma-1){(\mathbf{k'\cdot V})
\over\|\mathbf{V}\|^2}
+\gamma{(\mathbf{k'\cdot u'})\over c^2}\right)\right]\cdot\mathbf{r} \nonumber\\ 
&-&\gamma\mathbf{k'}\cdot(\mathbf{u'}+\mathbf{V})t.
\end{eqnarray}
If the frames $S$ and $S'$ are in standard configuration, then $\mathbf{V}=(V,0,0)$, $\mathbf{k}=(k_x,k_y,k_z)$, 
$\mathbf{k'}=(k'_x,k'_y,k'_z)$, $\mathbf{u}=(u_x,u_y,u_z)$, $\mathbf{u'}=(u'_x,u'_y,u'_z)$, and
$\mathbf{r}=(x,y,z)$. Substituting in Eq. (8) and comparing the terms for arbitrary $x,y,z,$ and $t$,
we obtain:
\begin{eqnarray}
&&k_x=\gamma\left(k'_x+{V\over c^2}(k'_x u'_x+k'_y u'_y+k'_z u'_z)\right),\\
&&k_y=k'_y,\\
&&k_z=k'_z,\\
&&k_x u_x + k_y u_y + k_z u_z=\nonumber\\
&&=\gamma(k'_x V + k'_x u'_x + k'_y u'_y + k'_z u'_z).
\end{eqnarray} 
The angular frequencies of the wave in the corresponding frames are
given by:
\begin{eqnarray}
\omega'&=&|\mathbf{k'\cdot u'}|=|k'_x u'_x + k'_y u'_y + k'_z u'_z|,\\
\omega&=&|\mathbf{k\cdot u}|=|k_x u_x + k_y u_y + k_z u_z|.
\end{eqnarray} 
Furthermore, by putting Eqs. (9), (10) and (11) into Eq. (12), and comparing the terms for arbitrary $k'_x$, 
$k'_y$ and $k'_z$, we obtain:
\begin{eqnarray}
u_x={u'_x+V\over 1+u'_x V/c^2},\\
u_y={u'_y/\gamma\over 1+u'_x V/c^2},\\
u_z={u'_z/\gamma\over 1+u'_x V/c^2},
\end{eqnarray} 
which are the relativistic velocity transformation formulas for the wave. The set of equations 
(9)--(12) and (15)--(17) describe the relativistic transformation of the wave characteristics from $S'$ to $S$. 
The reverse transformation is accomplished via $V$-reversal, that is, by replacing $V$ with --$V$ 
and interchanging the primed and unprimed quantities. 

We will demonstrate the effect of relativistically-induced optical anisotropy by applying the above 
analysis to describe the transverse drag of light \cite{gjurchinovski1,gjurchinovski2}. 
In the rest frame $S'$ of the medium, a plane wave propagates in the direction of the positive $y'$-axis
at a velocity $\mathbf{u'}=(0,u',0)$ (see Fig. 2). 
\begin{figure}
\includegraphics[width=.40\textwidth,height=!]{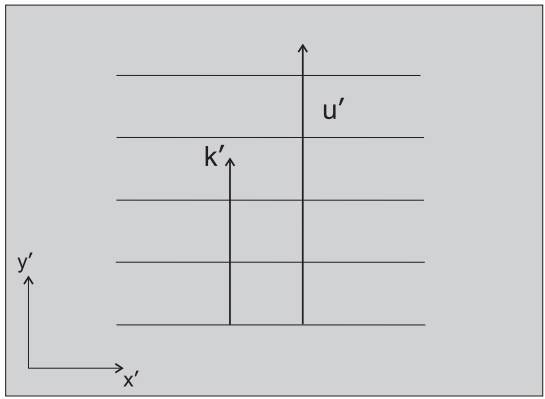}
\caption{A plane wave advancing in a stationary medium at a speed $u'$ in the direction of
the positive $y'$-axis.}
\end{figure} 
By assuming that the medium is homogeneous, isotropic and nondispersive in $S'$,
we have $\mathbf{k'}=(0,k',0)$. Evidently, the velocity of the wave and the wavefront normal in the 
$S'$-frame are parallel and unidirectional. Also, for the angular frequency of the wave in $S'$, we have 
$\omega'=|\mathbf{k'\cdot u'}|=k'u'$.
With respect to the $S$-frame in which the medium is moving at a 
velocity $\mathbf{V}=(V,0,0)$, from Eqs. (9)--(11) and Eqs. (15)--(17) we obtain $\mathbf{k}=(\gamma Vk'u'/c^2,k',0)$
and $\mathbf{u}=(V,u'/\gamma,0)$. Consequently, $\mathbf{k}\times\mathbf{u}=[0,0,k'V(u'^2/c^2-1)]$, which means
that the velocity of the wave and the wavefront normal in the $S$-frame are generally not parallel (see Fig. 3), 
except when $u'=c$, which is the vacuum case. 
Also, $\omega=|\mathbf{k\cdot u}|=\gamma\omega'$, which is the well-known formula for the transverse Doppler effect.
\begin{figure}
\includegraphics[width=.40\textwidth,height=!]{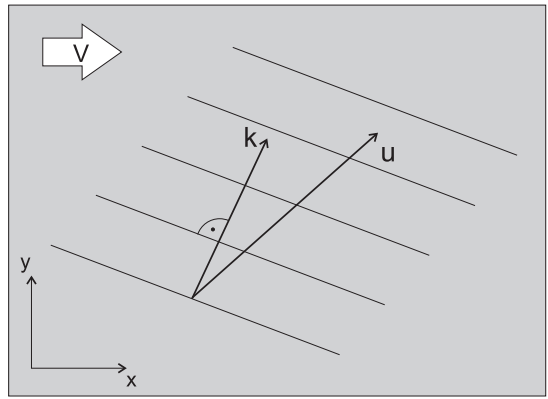}
\caption{A snapshot of the wave propagation in Fig. 2 with respect to the frame where
the medium is moving at a constant speed $V$ to the right. Notice that $\mathbf{k}$ and $\mathbf{u}$
are not parallel, which means that the light ray is not normal to the wavefronts.}
\end{figure} 

We conclude this section with the analysis of the longitudinal Fresnel-Fizeau light drag, 
which is the case discussed in the aforementioned EPL letter \cite{huang}. Now, in the rest frame $S'$ of 
the medium (Fig. 4), the wave  propagates in the direction of the positive $x'$-axis at a velocity 
$\mathbf{u'}=(u',0,0)$. 
\begin{figure}
\includegraphics[width=.40\textwidth,height=!]{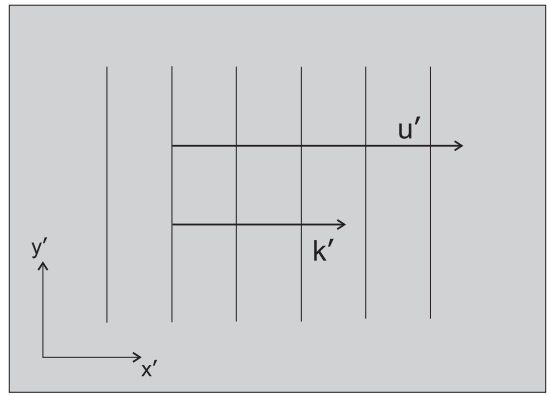}
\caption{A plane wave propagates in a stationary medium at a speed $u'$ in the direction of
the positive $x'$-axis.}
\end{figure} 
The wave vector describing
the wavefront normal is $\mathbf{k'}=(k',0,0)$, and the angular frequency of the wave is 
$\omega'=|\mathbf{k'\cdot u'}|=k'u'$. Transforming to the $S$ frame in which the medium moves at a 
velocity $\mathbf{V}=(V,0,0)$, we obtain $\mathbf{k}=[\gamma k'(1+Vu'/c^2),0,0]$ and 
$\mathbf{u}=[(u'+V)/(1+u'V/c^2),0,0]$. Consequently, the angular frequency of the wave in $S$-frame is
$\omega=|\mathbf{k\cdot u}|=\gamma|\omega'+k'V|$. If the medium moves in the negative $x$-axis, 
the wave characteristics are obtained by replacing $V$ with $-V$. Hence, 
$\mathbf{k}=[\gamma k'(1-Vu'/c^2),0,0]$, $\mathbf{u}=[(u'-V)/(1-u'V/c^2),0,0]$ and 
$\omega=|\mathbf{k\cdot u}|=\gamma|\omega'-k'V|$. In the case of ``superluminal'' motion of the medium,
$u'<V<c$, which implies $k_x>0$ and $u_x<0$. Hence, the dragging of the wave by the
medium is overwhelming, and the wave will propagate along the negative $x$-axis (see Fig. 5). 
\begin{figure}
\includegraphics[width=.40\textwidth,height=!]{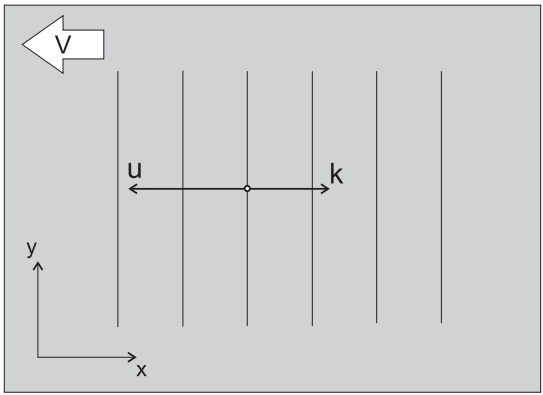}
\caption{A snapshot of the wave in Fig. 4 with respect to the frame where
the medium is moving at a ``superluminal'' speed $V$ to the left. In this case, 
although $\mathbf{k}$ and $\mathbf{u}$ are parallel, they are pointing in opposite directions.}
\end{figure} 
That the wave vector and the velocity of the wave in $S$-frame remain parallel, but in opposite directions,
is a robust example of the relativistically-induced optical anisotropy at work.
Nevertheless, the reader may notice that the frequency of the wave remains positive by definition,
in spite of the fact that $\mathbf{k\cdot u}<0$.

\section{Conclusion}

When analyzing the wave propagation in a medium with respect to different inertial reference
frames, one must take into consideration the effect of relativistically-induced optical
anisotropy due to the motion of the medium. In this sense, the angle between the wave vector 
(the wavefront normal) and the velocity of the wave is not Lorentz-invariant, and furthermore, 
the dot product between these vectors may have different signs with respect to different reference frames. 

To employ the concept of phase invariance of the wave among inertial frames  
one should use the correct expression for the wave four-vector $k^\mu$: 
\begin{equation} 
k^\mu=\left(\mathbf{k},{\mathbf{k\cdot u}\over c}\right),
\end{equation} 
where $\mathbf{k}$ is the wave three-vector, and $\mathbf{u}$ is the velocity 
of the wave. Employing the less-general, but widely used expression for the four-vector
in the form $k^\mu=(\mathbf{k},\omega/c)$ as in the recent EPL letter, one may
be tempted into a spurious conclusion that the invariance of the phase of waves among 
inertial frames is questionable.

\end{document}